# Cooperative Peer-to-Peer Electricity Trading of Nanogrid Clusters for Decentralized Power Management Based on Predictions of Load Demand and PV Power Generation

Sangkeum Lee[a], Hojun Jin[a], Luiz Felipe Vecchietti[a], Junhee Hong[b], Ki-Bum Park[a], and Dongsoo Har[a*]

[a] The Cho Chun Shik Graduate School of Green Transportation, Korea Advanced Institute of Science and Technology, Daejeon 34141, South Korea (e-mail: {sd974201, hjjin1995, lfelipesv, ki-bum.park, dshar}@kaist.ac.kr).

[b] The Department of Energy IT, Gachon University, 1342 Seongnam-daero, Seongnam, South Korea (e-mail: hongpa@gachon.ac.kr).

* Corresponding author at: The Cho Chun Shik Graduate School of Green Transportation, KAIST,291 Daehak-ro, Daejeon, Republic of Korea. Tel: +82 42 350 1267. E-mail address: dshar@kaist.ac.kr.

*Abstract*

This paper presents the power management of the nanogrid clusters assisted by a novel peer-to-peer(P2P) electricity trading. In our work, unbalance of power consumption among clusters is mitigated by the proposed P2P trading method. For power management of individual clusters, multi-objective optimization simultaneously minimizing total power consumption, portion of grid power consumption, and total delay incurred by scheduling is attempted. A renewable power source photovoltaic(PV) system is adopted for each cluster as a secondary source. The temporal surplus of self-supply PV power of a cluster can be sold through P2P trading to another cluster (s) experiencing temporal power shortage. The cluster in temporal shortage of electric power buys the PV power to reduce peak load and total delay. In P2P trading, a cooperative game model is used for buyers and sellers to maximize their welfare. To increase P2P trading efficiency, future trends of load demand and PV power production are considered for power management of each cluster to resolve instantaneous unbalance between load demand and PV power production. To this end, a gated recurrent unit network is used to forecast future load demand and future PV power production. Simulations verify the effectiveness of the proposed P2P trading for nanogrid clusters.

I.  INTRODUCTION

Nanogrid is unique in that it is an islandable DC-centric power grid. Because it is digitally controlled, every electric load operated with digital control can be connected to nanogrid. Being islandable, nanogrid is implemented in combination with distributed energy resource (DER) systems such as photovoltaic (PV) systems, energy storage systems (ESSs), and electric vehicles (EVs). With these components in the nanogrid, more diverse power flow and subsequently improved power management can be realized. Since digital control of individual loads requires specific information pf them, digital interfaces like USB port and Power-over-Ethernet are adopted for communication of load information and power delivery. The terms "load" and "appliance" are interchangeably used. With the specifications of electric appliances, power management at the appliance level [1] can be performed. Unlike typical power management attempting to establish system-level policy with a large scale power grid, power management of nanogrid involves peak load management in peak hours and operation scheduling of individual appliances. Electric loads allowing scheduling are flexible loads and other electric loads in immediate demand are non-flexible loads. From this perspective, previous works [2],[3] investigated optimal power management of nanogrid to simultaneously minimize total power consumption and total (accumulated) delay of electric loads. In [2],[3], temporal use of the specific electric appliance and a Markov chain(MC) model describing resident mobility are considered as resident behavior within the residence. In [1]-[4], ESS, PV source, and EV battery are considered secondary power sources and taken into account for the overall power management of nanogrid. These secondary power sources supply additional electricity when the instantaneous power consumption of nanogrid exceeds the maximum allowed power consumption.

P2P trading is the buying and selling of electricity between two or more parties connected by a power grid. When it is concerned with DER systems, it reduces the unbalanced distribution of power consumption among involved parties [4]. Since each nanogrid has its smart meter, nanogrids can share individual real-time power statements through the P2P network. Nanogrid cluster that is considered an essential structure for P2P trading consists of multiple grid-connected nanogrids and uses DER systems of the nanogrids for its power management. When a cluster has excess energy or shortage of energy, the cluster acts as a seller(producer) or a consumer. A cluster that acts as a producer and a consumer alternately over time is called a prosumer. By the P2P trading, a cluster's surplus PV power is used to reduce peak load, grid electricity cost, and delay of load scheduling of other cluster(s) [5]. Various architectures for P2P trading have been introduced in the literature. The P2P trading architectures with DER systems can be classified into local, centralized, decentralized, distributed, and hierarchical ones [6],[7] . In local P2P trading architecture, each party(peer) participating in P2P trading is allowed to determine which peer to trade electricity, according to trading objectives [8]. Local P2P trading can be implemented in the form of an online service. The main advantage of local P2P trading is the availability of a fail-safe design against communication disturbance [9]. In [10], decentralized P2P trading taking into account network issues such as overvoltage is presented. Decentralized P2P trading is in general more robust and more scalable in transactions than a centralized one whereas the centralized one is more cost-efficient. Centralized P2P trading is typically composed of a single controller for the entire grid network with communication links. The central controller enables higher transaction speed in P2P trading [11]. In a centralized market for electricity trading, an energy allocation algorithm is developed to achieve the market equilibrium based on all parties' information, which in effect maximizes the global welfare [12]. In distributed P2P trading, however, agents(participating parties) communicate through multiple local agents [13] and submit offers and bids considering their preferences and costs. In [14], the hierarchical control architecture for P2P trading is proposed for a tree-shaped command/feedback control system based on multiple layers. In this work, P2P trading is applied to decentralized networks for power management of cooperative nanogrid clusters sharing welfare.

For the P2P trading system, a 3-layer architecture consisting of power network, information network, and business network is presented in [15]. Electric devices in power network are classified into 1) loads, e.g., flexible or non-flexible, and 2) storage, e.g., batteries of ESS and EV, and 3) power production, e.g., PV system, wind turbines, combined heat and power system [8]. PV system can be used in a stand-alone configuration without ESS for P2P electricity trading. EVs are becoming more popular due to their environment-friendly operation and capacities in assisting the grid via vehicle-to-grid technologies [16],[17]. From the perspective of the information network, the P2P trading system should reliably and securely manage the electricity trading between peers. The information obtained by the smart meters can be managed securely by security technologies such as blockchain [18], [19] . Recent articles such as [20], [21], [22] survey security issues in grid system and their solutions. Privacy and security of smart meter are investigated in [23],[24]. The information required for P2P trading can be shared and aggregated among smart meters by connecting them according to specific protocols like IPv6 over low-power wireless personal area networks [25]. Based on the shared information, the P2P trading system can facilitate electricity trading, considering the time and location of power production, storage, and power consumption [26]. The rate of traded electricity is typically determined by the system marginal price (SMP) and the renewable energy certificate (REC) issued by authorities like the Korea Power Exchange (KPX) [27].

Game theory is the study of mathematical models to analyze strategic interaction among rational decision-makers where the outcome from a player's action is dependent on the actions of other players [28], and is widely used for electricity trading in the open market. Game theory can be applied to the non-cooperative game and cooperative game [28]. Each prosumer attempts to maximize its profit, and all prosumers finally come to an equilibrium state [4]. In [29], a non-cooperative game is presented, implementing electricity trading with a set of energy storage units to maximize the profit while taking into account the trading price. Nash game, one of the non-cooperative games, has been applied to maximize profits of the utility companies when the consumers want to maximize their welfare [30]. Stackelberg's leadership model that is a strategic game theory in economics in which the leader firm moves first and then the follower firms move sequentially is applied to a distributed mechanism for electricity trading in the smart grid [29]. By establishing a Stackelberg leadership model for producers, the producers are required to behave as leaders to maximize their profits and consumers are required to act as followers to maximize their welfare [31]. On the contrary, a cooperative game seeks to provide incentives to independent decision-makers when acting together as one entity. The cooperative game model is developed to incentivize prosumers to form coalitions in P2P energy trading [32]. In [33], a cooperative power dispatching algorithm is applied for microgrids to minimize network operation costs and to satisfy stochastic demands within the grid. Cooperative P2P trading based on Shapley value that deals with dividing the surplus among self-interested agents in a coalition is presented in [34] for the efficient use of energy in remote communities. In the long term, the cooperative game framework is more effective for increasing energy efficiency and facilitating the local power consumption of DERs.

Artificial neural networks(ANNs) have been increasingly used for research works related to smart grids [35]. Convolutional neural network (CNN) architecture fit to pattern recognition is used to predict energy demand in a real-world smart grid [36]. Recurrent neural network (RNN) fit for processing of time series data is extensively utilized in the electricity load prediction [37], energy market prediction [38], and renewable energy prediction [39]. Dynamic programming and deep RNN are used to get optimal real-time scheduling policy in microgrid [40]. A feed-forward neural network is applied to PV systems for solar irradiance forecast over 24 hours [41]. In [42], the PV power forecasting model is proposed using ANNs. In [43] and [44], a gated recurrent unit (GRU) network, a variant of the RNN, is used for forecasting the short-term trend of PV power production. Other applications of ANN for renewable energy sources are reviewed in [45], [46]. Load variation of the smart grid affects the electricity trading process, market share, and overall profits in the electricity market [47], [48]. To forecast electricity load in microgrids, a self-recurrent wavelet neural network is utilized in [49]. In [50], the GRU network is used to predict electrical load. The GRU-CNN hybrid neural network is utilized for short-term load forecasting. Feature vector of time series data for short-term load forecasting is extracted by GRU network and feature vector of other high-dimensional data is extracted by CNN [51]. In our work, the GRU network is used for forecasting both trends of load demand and PV power production, considering the influence of historical trends of load demand and PV power production on the future trends of them. While PV power production is on supply side of power management, load variation is on demand side of power management.

In this paper, the power management of the nanogrid cluster with the help of P2P trading is presented. Each cluster takes two electricity sources: utility grid and PV system. The load demand of the cluster consists of flexible and non-flexible loads. EVs are included as flexible loads and heater+ventilation fan+air conditioner (HVAC) for sustained living comfort is categorized into non-flexible loads. From here, "flexible loads" indicate flexible loads including EVs and "non-flexible loads" represent non-flexible loads including HVAC. The temporal use of electric loads by the resident is determined by resident behavior. Resident behavior is described by resident mobility modeled by the MC. EV charging is initiated and scheduled depending on the charging probability model provided by the statistics of EVs registered in Jeju Island, Korea [52]. To utilize the PV power remaining after supplying for residents' load demand in cluster, P2P electricity trading is executed between clusters. When deciding the role of each cluster in P2P trading, future load demand and future PV power production, both of which are forecasted by GRU network, are considered within the framework of multi-objective optimization simultaneously minimizing total power consumption, total grid power consumption, and total delay. It is noted that the operation of HVAC is controlled according to an individual service plan. Once the role of each cluster is determined, P2P trading is executed in a cooperative game model. With traded PV power, eventual power management is performed. When self-supplied PV power and PV power bought in P2P trading are insufficient for instantaneous power consumption of a cluster, grid power is used. The contributions of this paper are as follows.

1. Power management of nanogrid cluster with the help of the proposed P2P trading method is presented. Renewable energy resource PV power is used as secondary power used ahead of primary grid power. Unlike conventional P2P trading based on instantaneous power consumption profile in relation to current PV power production, the proposed P2P trading method is based on current and forecasted load demand and PV power production.

2. Cost-efficient P2P trading method is proposed. The use of the GRU network for P2P trading between clusters leads to reduced electricity cost and total delay in the scheduling of flexible loads. P2P electricity trading based on future trends of load demand and PV power production is more cost-efficient than the P2P electricity trading solely based on instantaneous variations of load demand and PV power production. To achieve reduction of electricity cost, demand response(DR) program as well as SMP is considered for power management of cluster

3. P2P trading conditions in terms of future load demand and PV power production are specified to assign the role of each cluster as a producer or a consumer. Based on the proposed P2P trading conditions, the amount of PV power for P2P electricity trading is determined via a cooperative game model. The electricity is traded with the objective to achieve cooperative power management between clusters.

4. Temporal variations of power consumption and total delay of nanogrid cluster assisted by P2P trading are investigated. Multi-objective optimization fit to power management of nanogrid cluster with P2P trading of PV power is proposed in appliance level rather than grid level typically dealt with power management of microgrid. Comparison of results of power management based on proposed P2P trading of PV power with results of power management obtained with conventional P2P trading shows a significant difference of total power consumption and total delay of the cluster.

Power resources consist of grid power and PV power. The goal of power management for nanogrid cluster is to reduce daily electricity cost quantified by DR program and SMP. Daily electricity cost is evaluated with grid power cost plus incremental or decremental cost of PV power traded. To reduce daily electricity cost, multi-objective optimization is attempted, taking short-term forecasted trends of load demand and PV power production into consideration. PV power is used as a secondary power resource and thus use of it affects reduction of electricity cost and total delay significantly. With EV being charged, grid power and PV power are consumed for EV charging and thus proper scheduling of EV charging is essential for power management. The operation of HVAC depends on the control mode or service plan according to the resident location.

The ESS mitigating intermittency of PV source is not used for this work because the use of ESS causes significantly increased SMP and thus weakens the cost advantage of PV power in comparison with grid power [1]. This paper is organized as follows. In Section II, the operation of nanogrid cluster with P2P trading are explained. Section III provides details of power management based on the proposed P2P trading method. Section IV presents simulation results obtained with 6 different settings of P2P electricity trading and Section V concludes this paper.

## II. OPERATION OF NANOGRID CLUSTER WITH P2P TRADING

*A. P2P trading system*

In Fig.1, the P2P electricity trading system is illustrated. It consists of 6 clusters, each composed of 3 nanogrids, and all clusters are electrically connected to each other for P2P trading. A PV system represented by rooftop PV panels of 3 nanogrids and their associated electronics is operated for the cluster. The smart meter monitors, records, and transmits the information on load demand and PV power production. Smart meters communicate with each other for smart contracts of P2P trading [53]. Data collected by smart meters is a source of information and is expected to be used in combination with other sources of information in the nanogrid.

The power packet transmission model is used for P2P trading. Utilization of PV power is optimized by cooperative electricity trading between producers and consumers. The producer or consumer's role is assigned to each cluster, as described in [54], in four steps: registration, routing, scheduling, and transmission steps. In the registration step, the cluster can be registered as a producer or a consumer. A power packet from producer would be transmitted to the consumer linked to an adjacent router. Hence, in the routing step, the trading controllers determine the consumer and the intermediate routers pass the consumer's power packet.

TABLE 1. Electric appliances placed in rooms of a house. 3 HVAC appliances are installed in every room of the house. Power rating of each electric appliance is also listed [1].

| Index | Kind (room number) | Power rating | Index | Kind (room number) | Power rating |
|---|---|---|---|---|---|
| 1 (HVAC, non-flexible) | Air-conditioner (1,2,3,4) | 1.2kW | 7 (flexible) | Washing machine (2) | 242W |
| 2 (HVAC, non-flexible) | Electric fan (1,2,3,4) | 60W | 8 (flexible) | Vacuum cleaner (3) | 1.07kW |
| 3 (HVAC, non-flexible) | Heater (1,2,3,4) | 1.16kW | 9 (flexible) | Iron (1) | 1.23kW |
| 4 (non-flexible) | Computer (3) | 255W | 10 (flexible) | Microwave oven (4) | 1.04kW |
| 5 (non-flexible) | TV (1) | 130W | 11 (flexible) | Rice cooker (4) | 1.03kW |
| 6 (flexible) | Audio (3) | 50W | 12 (flexible) | Hair dryer (2) | 1kW |

In this paper, the cooperative game model is applied to P2P trading. The cooperative game model focuses on how independent clusters act together as one entity to maximize global welfare in the game [16]. As shown in Fig.2, P2P trading occurs in two cases; when PV power supplied by producers is larger than power demand by consumers and when PV power supplied by producers is less than power demand by consumers. If supplied PV power is sufficient for the demand, the amount of power demand is set as the amount of power for P2P trading. If supplied PV power is insufficient for P2P trading, the supplied PV power is set as the amount of power for P2P trading. Power supplied for P2P trading by each producer is proportional to the amount of excess power each producer has, and each power demanded to P2P trading by each consumer is also proportional to the amount of power each consumer requires.

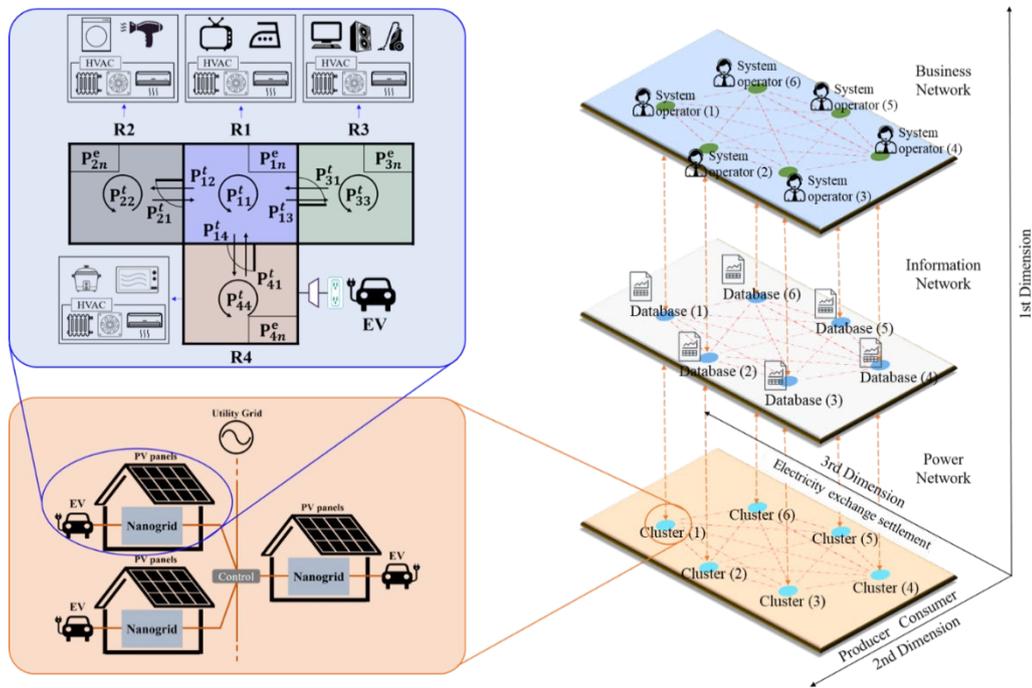

Fig.1. Architecture of P2P electricity trading system [15].

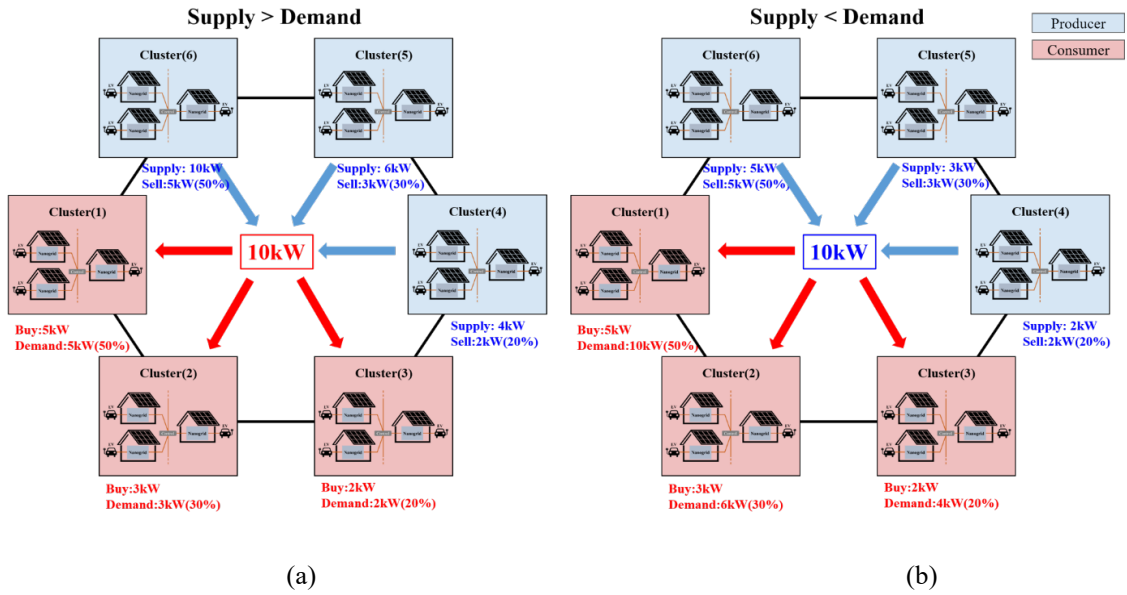

Fig.2. Cooperative P2P trading: (a) when supply is greater than demand; (b) when supply is less than demand.

A 3-network system architecture shown in Fig.1 is considered for P2P trading. There are 3 dimensions of P2P trading. The 1st dimension is related to the decomposition of the P2P trading system into 3 interactive networks [15]: power network, information network, and business network. The power network comprises physical components of the power system, such as feeders, smart meters, transformers, loads, and DERs. In the information network, communication devices, protocols, application program, and information flow are involved. For power management, each cluster needs to exchange the information on power demands, renewable energy generations, and power assisting devices of its own. Communication devices consist of sensors, wired/wireless communication connections, switches, routers, and servers [8]. Communication program can be utilized in information transfer and file exchange between clusters. This network mainly involves consumers, producers, distribution system operators (DSOs), and electricity market regulators.

The 2nd dimension of the P2P trading system is about the role of peers, e.g., producer or consumer. Each cluster can be either a producer or a consumer in inter-cluster electricity transaction depending on load demand in relation to PV power production

The 3rd dimension indicates electricity exchange settlement between clusters. It is sequentially implemented by bidding, electricity exchanging, and settlement. In the bidding process, customers, e.g., producers, consumers, and DSOs and agree to participate in P2P trading for power management of the overall system. Based on the bidding process, the generated energy is distributed to other peers through electricity exchanging process.

*B. Structure of nanogrid cluster*

Each nanogrid installed in a house of 4 rooms is composed of flexible and non-flexible loads, a PV system, and the utility grid. List of flexible loads and non-flexible loads with their power ratings and installed room indices can be found in TABLE 1 adopted from [1]. Flexible loads can be scheduled to another time interval, e.g., 10 minute long time slot in our work, to reduce peak load within the maximum allowed delay. Non-flexible loads are operated immediately on demand. One resident per house is assumed and the resident moves to other room or stays in every 10 minutes in the same room according to transition probabilities $P_{ij}^t$, where *i* is the index of room before transition and *j* is the index of room after transition. Resident mobility describing random transition between rooms is modeled by a MC model that the probability of transition from previous room to current room depends only on previous room. Transition probability $P_{ij}^t$ is equally assigned for all *j* indices. In case of room 1(R1) in Fig.1, random transition probability to other room or same room is 1/4 whereas it is 1/3 for other rooms. The probabilistic use of each electric appliance over time is determined by the Korean Time Use Survey (KTUS) data [55] obtained by the KPX with 500 residences. The specific emission probability of an appliance is determined in [1] as follows. The charging probability of an EV, which indicates the probability of initiation of charging in each hour, is shown in Fig.3. The charging probability is obtained from 6,080 EVs registered in Jeju Island, Korea.

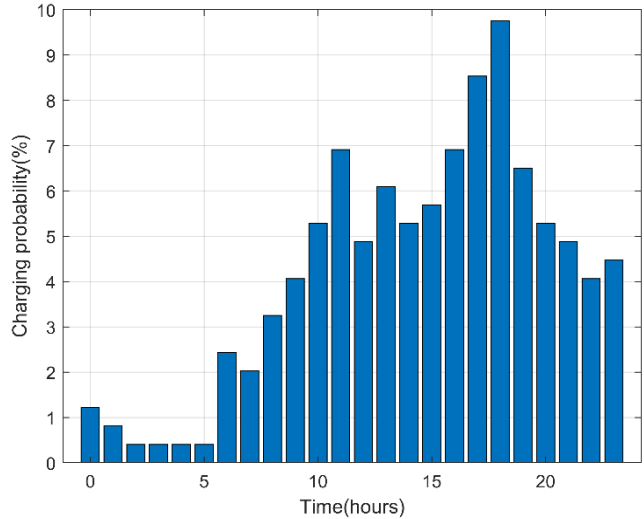

Fig.3. Charging probability (%) of an EV over a day.

*C. Rate of electricity according to DR program and SMP*

DR program is utilized to offer incentives to customers that use less electric power during peak hours. Depending on the DR program, the residential loads, particularly flexible loads, can be scheduled to reduce the grid electricity cost. The DR program offers a lot of potential merits, such as cost and emission reduction of the power plant, reliability improvement in power system, and reduced dependence on overseas fuels [46]. Rate of electricity by DR program initiated by Korea Electric Power Corporation (KEPCO) varies $0.05/kWh over time zone 23:00～09:00, $0.1/kWh over time zones 09:00～10:00, 12:00～13:00, 17:00～23:00, and $0.18/kWh over time zones 10:00～12:00, 13:00～17:00, as illustrated in Fig.4(a). Through P2P trading, the power generated by PV system is sold to other clusters. Figure 4(b) shows the daily PV power production of a PV system averaged over each month of a PV system throughout the day [27]. In P2P trading, the purchase/sale price of the power generated by the PV system is determined by

$$B_{PV}(n) = [PW_{PV,P2P}(n)/6] * SMP(n) \qquad (1)$$

where $B_{PV}(n)$ is the electricity cost due to the consumption of PV energy generated by PV system at the *n*-th time interval and $PW_{PV,P2P}(n)$ is the PV power traded in P2P trading and $SMP(n)$ is the SMP price at the *n*-th time interval and the factor (1/6) represents 10 minutes in the unit of hour. In simulations, a single value of $PW_{PV,P2P}(n)$ is used for the *n*-th time interval.

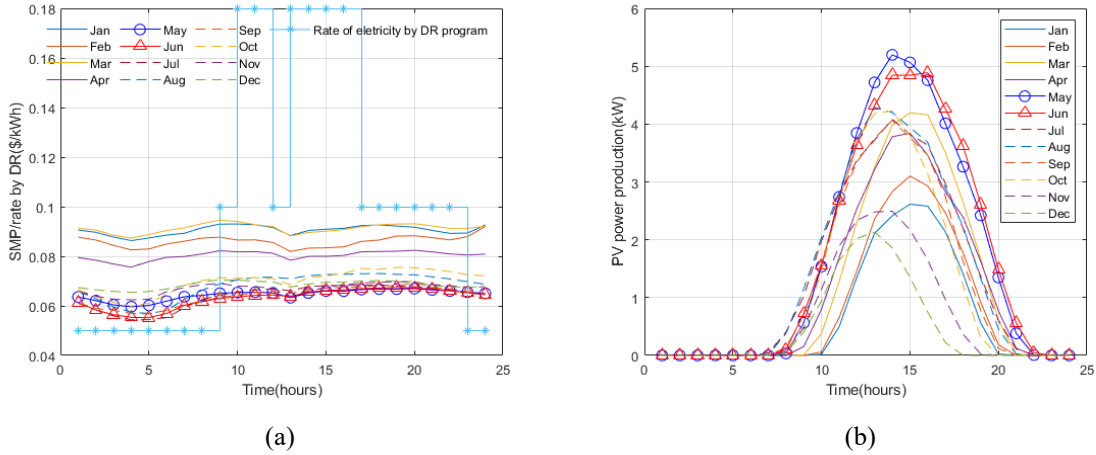

Fig.4. Variation of PV power production and SMP: (a) rate of electricity by DR program and average monthly SMP; (b) daily PV power production averaged over each month.

*D. Prediction of load demand and PV power production for P2P trading*

When the initial supply-demand relationship among clusters is established, the opportunity to join P2P trading is offered to clusters. In Fig.5(a), all clusters are connected to the utility grid through a point of common coupling(PCC). The utility grid supplies the necessary power whenever the local PV power production is insufficient to meet the local load demand. On the other hand, when PV power is sufficient to meet the load demand in a cluster, the remaining PV power can be sold by P2P electricity trading. The market operator for P2P trading is accountable for a P2P matching relationship based on a cooperative game model.

Long short-term memory (LSTM) architecture is proposed to prevent the vanishing or exploding gradient problems. To reduce computational time and cost of LSTM, the GRU network is introduced later. Similarly to the LSTM cell, the GRU cell controls the information flow without using a memory unit since it reveals the hidden contents without any control. In this paper, the GRU network is used to forecast load demand and PV power production of each cluster. The GRU network consists of 6 GRU layers and 3 fully connected layers as presented in Fig.5(b). The dropout of fully connected layers is to prevent the overfitting of the GRU network. The number of GRU layers and the number of input data to the input GRU layer are determined by the trial-and-error method like other works [56]. The number of inputs and outputs of the GRU network are determined through the evaluation of root-mean-squared-error(RMSE) of forecasted load demand and PV power production.

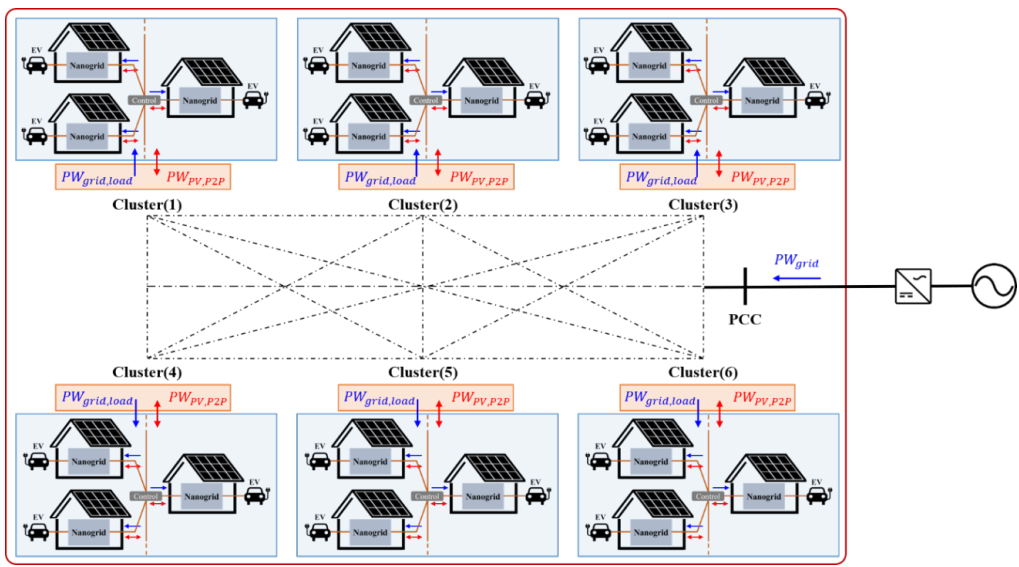

(a)

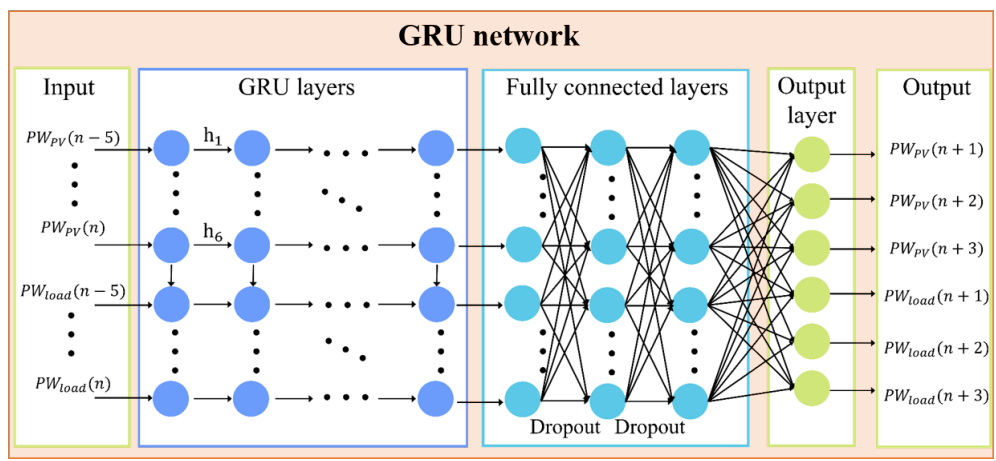

(b)

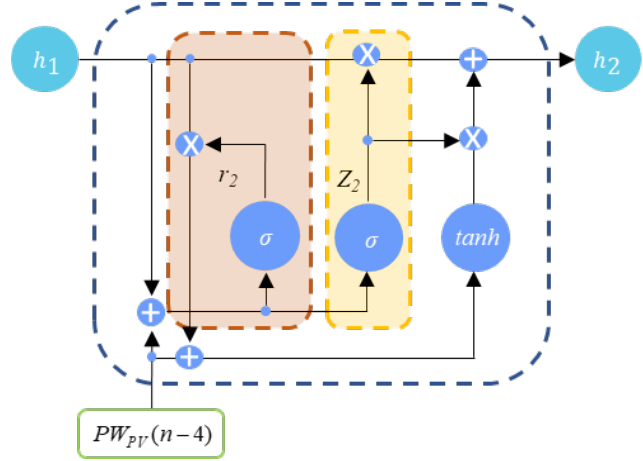

(c)

Fig.5. Grid-connected clusters and GRU network for power management of clusters: (a) grid-connected clusters considered in P2P trading; (b) structure of GRU network with past inputs and future outputs; (c) GRU cell.

The 1 year dataset of PV power production and the 1 year dataset of outdoor temperature synchronized with the PV power production are obtained from the Korea Meteorological Administration. PV power production curve of the dataset takes 2kW peak power around 13hrs throughout 1 year. As suggested in [57], trend of outdoor temperature variation is well matched with the trend of PV power production. Since HVAC operation is fundamentally affected by outdoor temperature variation, training with a historical dataset of HVAC operation represents training with outdoor temperature variation. Therefore, the temporal power consumption of HVAC is implicitly synchronized with the trend of PV power production. The dataset of load demand for 1 year is obtained from the cluster's power management described in Section III. Each dataset is divided into 80 percent of training set and 20 percent of validation set. The number of outputs of load demand and PV power production is relatively small due to our interest in short-term forecasting of load demand and PV power production. When the equal number of past inputs of load demand and PV power production is 6 and the equal number of future outputs of load demand and PV power production is 3, the RMSE obtained during validation of GRU network is about 10%, ranging between 7% and 14%, and is among the lowest. The inputs of GRU network are load demand $PW_{load}(n-i)$ and PV power production $PW_{PV}(n-i)$ at the (n-i)-th time interval, where i=0,…,5 and the outputs of GRU network are load demand $PW_{load}(n+j)$ and PV power production $PW_{PV}(n+j)$ at the (n+j)-th time interval, where j=1,…,3. As shown in Fig.5(c), a GRU cell **[58]** contains two control gates, the update gate centered around the coefficient $z_2$, the reset gate around $r_2$, and the input vector of the GRU cell $PW_{PV}(n-4)$. The update gate controls how much the state information $h_1$ at the previous time step is delivered to the current time step. The reset gate is used from the GRU cell to decide how much of the past information is to forget [59], [58].

### III. POWER MANAGEMENT OF NANOGRID CLUSTER WITH P2P ELECTRICITY TRADING

This section describes the details of the power management of nanogrid cluster with P2P electricity trading. The smart meter of the cluster is assumed that it can control operation of all loads and devices. The smart meter is used to capture raw voltage and current signals separately, and the power consumption of a cluster can be computed from the discrete signals. Prior to multi-objective optimization, the role of each cluster is determined. To this end, trade mode buy or sell is assigned as a value to the switching function for P2P trading. The value of the switching function is used for multi-objective optimization attempting to minimize peak load, grid dependency, and total delay of flexible appliances simultaneously.

#### A. Proposed P2P trading method with clusters

P2P trading can be described by switching functions. The switching function $O_{PV,P2P}(n)$ for conventional P2P trading is given according to emergency condition involving the maximum allowed power consumption $PW^{max}$ as follows

$$O_{PV,P2P}(n) = \begin{cases} +1(buy) & if\ (PW_{load}(n) - PW_{PV}(n) > PW^{max}) \\ -1(sell) & if\ (PW_{load}(n) - PW_{PV}(n) < PW^{max}\ and\ PW_{PV}(n) > 0) \end{cases} \quad (2a)$$

where $PW_{load}(n)$, $PW_{PV}(n)$ are power consumption of all loads being used, self-supply PV power, respectively, at the $n$-th time interval. The $PW_{load}(n)$ is obtained as the sum of individual power consumption of flexible loads and non-flexible loads in the cluster at the $n$-th time interval, without scheduling of flexible loads. In the conventional P2P trading method [60] in (2a), when the difference between load demand and PV power production at current time interval $PW_{load}(n) - PW_{PV}(n)$ is bigger than $PW^{max}$, $O_{PV,P2P}(n)$ becomes +1(buy). If it is smaller with available $PW_{PV}(n)$, $O_{PV,P2P}(n)$ becomes -1(sell). On the other hand, buy or sell of PV power is determined by proposed method as follows

$$O_{PV,P2P}(n) = \begin{cases} +1(buy) & if\ (\sum_{k=0}^{K}(PW_{load}(n+k) - PW_{PV}(n+k)) > (K+1)PW^{max}) \\ -1(sell) & if\ (\sum_{k=0}^{K}(PW_{load}(n+k) - PW_{PV}(n+k)) < (K+1)PW^{max}\ and\ PW_{PV}(n) > 0) \end{cases} \quad (2b)$$

In the proposed P2P trading method, the current and future states of load demand and PV power production are taken into account for P2P trading. If $\sum_{k=0}^{K}(PW_{load}(n+k) - PW_{PV}(n+k))$, which indicates accumulated difference, is bigger than $(K+1)PW^{max}$, $O_{PV,P2P}(n)$ becomes +1(buy). If it is smaller than $(K+1)PW^{max}$ and $PW_{PV}(n)$ is positive, $O_{PV,P2P}(n)$ becomes -1(sell). Note that practically impossible equality corresponding to "idle" mode is omitted in (2a), (2b) for brevity. Future states of load demand and PV power production are predicted by GRU network described in subsection II.D after proper training of it. Once the role of each cluster in P2P trading is determined, PV power for P2P trading $PW_{PV,P2P}(n)$ as a producer is determined by the ratio of the PV power supplied by each producer to the total supplied PV power for P2P trading and $PW_{PV,P2P}(n)$ as a consumer is determined by the ratio of the PV power required by each consumer to the total required PV power for P2P trading, according to cooperative game model explained in Fig.2.

From (2a) for conventional P2P trading, the amount of PV power to buy is $PW_{load}(n) - PW_{PV}(n) - PW^{\max}(n)$ and the amount of PV power to sell is $-PW_{load}(n) + PW_{PV}(n)$. On the other hand, from (2b) for proposed P2P trading, the amount of PV power to buy is $\sum_{k=0}^{K}(PW_{load}(n+k) - PW_{PV}(n+k) - PW^{\max})$ and the amount of PV power to sell is $\sum_{k=0}^{K}(-PW_{load}(n+k) + PW_{PV}(n+k))$. Once the amount of PV power to sell or buy is determined by cooperative game model as $PW_{PV,P2P}(n)$, the $PW_{PV,P2P}(n)$ is plugged in multi-objective optimization framework as follows.

B. *Multi-objective optimization for power management of a cluster*

   1) *Minimization of peak load(electricity cost) in a cluster*

The first objective function to minimize by scheduling of flexible loads is peak load. The peak load shifting is achieved by the scheduling of flexible loads and P2P trading as follows

$$\min_{(O_{flex,1},...,O_{flex,I})} \left[ \left( \sum_{i=1}^{I} PW_{flex,i}(n) * O_{flex,i}(n) + \sum_{k=1}^{K} PW_{non-flex,k}(n) * O_{non-flex,k}(n) \right) * \left( \frac{\frac{1}{6} * \left( \underbrace{PW_{grid,load}(n) * O_{grid,load}(n) * EC(n)}_{w/o\,P2P\,trading} + PW_{PV,P2P}(n) * O_{PV,P2P}(n) * SMP(n) \right)}{\underbrace{PW_{grid,load}(n) * O_{grid,load}(n) + PW_{PV,load}(n) * O_{PV,load}(n)}_{w/o\,P2P\,trading} + PW_{PV,P2P}(n) * O_{PV,P2P}(n)} \right) \right]_{w/\,P2P\,trading} \quad (3)$$

where $PW_{flex,i}(n)$ is the power consumption of the *i*-th flexible load and $PW_{non-flex,k}(n)$ is the power consumption of the *k*-th non-flexible load and $PW_{PV,P2P}(n)$ is the PV power sold/bought in P2P trading between clusters. The $O_{flex,i}(n)$ and $O_{non-flex,k}(n)$ are the switching functions of the *i*-th flexible load and the *k*-th non-flexible load, respectively and $O_{PV,P2P}(n)$ is the switching function of P2P trading for the cluster and $EC(n)$ is the rate of grid electricity. $PW_{grid,load}(n)$ is the grid power consumption due to load demand and $O_{grid,load}(n)$ is the switching function of the $PW_{grid,load}(n)$ and $PW_{PV,load}(n)$ is the PV power consumption and $O_{PV,load}(n)$ is the switching function of the $PW_{PV,load}(n)$ and $PW_{PV,P2P}(n)$ is the consumption of traded PV power and $O_{PV,P2P}(n)$ is the switching function of the $PW_{PV,P2P}(n)$. These switching functions, except $O_{PV,P2P}(n)$, take values 0 or 1. The factor (1/6) represents 10 minutes(=1/6 hr). The unit of the objective function is electricity cost($) over 10 minutes. The $PW_{grid,load}(n)$ is consumed by non-flexible loads and flexible loads. For convenience, $PW_{PV,load}(n)$ is used for non-flexible loads prior to the use for flexible loads and fractional power consumption of some load is PV power consumption and the remaining fraction is grid power consumption. Self-supply PV power is used prior to traded PV power. The total electricity cost consisting of individual costs of electric loads should be minimized with respect to flexible load scheduling by controlling the switching functions $O_{flex,1} \ldots O_{flex,I}$ of flexible loads. On the contrary, the switching functions of non-flexible loads cannot be controlled. The term "w/ P2P trading" involves the PV power variable related to P2P trading. The self-supply PV power is not taken into account for electricity cost, since it is considered free. As shown in (3), the electricity cost is evaluated in a weighted structure. The portion of electricity cost is due to grid power consumption and remaining portion of electricity cost is due to the consumption of traded PV power.

*2) Minimization of grid dependency with each cluster*

The self-supply capacity of the RES becomes more important for eco-friendly operation of nanogrid cluster. Even when it is more economical to consume grid power at low rate regulated by DR program, PV power is promoted for the eco-friendly operation. This motivation has profound effects on the reduction of fossil energy and enhancement of energy efficiency as more energy is consumed locally instead of delivered via long transmission lines[1]. Minimization in (3) is attempted in conjunction with the decrement of power consumption supplied by the utility grid and increment of the PV power by equal amount. Reduction of power consumption supplied by the utility grid as much as possible can be achieved from minimization as follows

$$\min_{\left(\substack{PW_{grid,load} \\ ,PW_{PV,load}}\right)} \left[ \underbrace{\underbrace{PW_{grid,load}(n)*O_{grid,load}(n) - PW_{PV,load}(n)*O_{PV,load}(n)}_{w/o\,P2P\,trading} - PW_{PV,P2P}(n)*O_{PV,P2P}(n)}_{w/\,P2P\,trading} \right]$$

(4a)

subject to

$$PW_{PV,load}(n) \leq PW_{PV}(n) \tag{4b}$$

$$PW_{grid,load}(n) < PW^{\max} \tag{4c}$$

The term "w/o P2P trading" contains power variables related to the utility grid and self-supply PV system. To reduce the dependency of grid power, the self-supply PV power is preferentially used for load until $PW_{PV,load}(n) = PW_{PV}(n)$, as suggested in (4b). Therefore, portion of power consumption of a load can be supplied by PV system and remaining portion is supplied by grid. The term "w/ P2P trading" includes power sold/bought in P2P trading. Grid power consumption of loads is regulated by the maximum power consumption constraint (4c). In order to reduce grid power consumption, the objective function in (4a) is to be minimized to ensure the use of P2P traded PV power $PW_{PV,P2P}(n)$ regardless of the rate of grid electricity regulated by the DR program.

*3) Minimization of total delay in each cluster*

To minimize load demand during peak hours, efficient scheduling of flexible loads is essential. Scheduling of flexible loads may incur inconvenience to residents when excessive delay occurs in using flexible loads. Therefore, minimizing the delay due to scheduling of flexible loads is significant for enhancing living convenience. Minimization of total delay of cluster can be achieved as follows

$$\min_{\left(O_{flex,1},...,O_{flex,I}\right)} \sum_{i=1}^{I} d_{flex,i}(n) * O_{flex,i}(n) \tag{5a}$$

subject to

$$d_{flex,i}(n) \leq d_{max} \quad i = 1,..,I \tag{5b}$$

where $d_{flex,i}(n)$ is the accumulated delay in scheduling the *i*-th flexible load. The accumulated delay in scheduling the *i*-th flexible load is evaluated according to the number of requests for using the flexible load. When, for example, the flexible load is requested twice according to the emission probability, each delay of each request is added up to be the accumulated delay of the flexible load [61]. The (5b) specifies a constraint that the accumulated delay of the *i*-th flexible load should be less than or equal to $d_{max}$.

Operation of HVAC is determined by the outdoor temperature and $CO_2$ density in relation to temperature and $CO_2$ density of each room [62]. For the resident's comfort, the target temperature of each room can be determined by the resident location. The value of $O_{PV,P2P}(n)$ in (2a), (2b) is substituted for (3), (4a), (4b). The multi-objective optimization formulated in (3), (4a)-(4c), (5a)-(5b) is carried out by the genetic algorithm(GA)[63]. Parameters of the GA are adopted from [64],[65]. The parameters of the GA used for the multi-objective optimization are crossover probability 0.8, mutation probability 0.01, the maximum number of generations 100, and population size 100.

IV. SIMULATIONS AND RESULTS

*A. Simulation Set-up*

System architecture presented in Fig.5(a) is considered in simulations. Utility grid and PV systems are taken as power sources. When self-supply PV power is insufficient for load demand, grid power or PV power obtained from P2P trading is used. For each house, a single resident is assumed living and load demand is generated according to resident behavior. The list of all loads except EVs is presented in TABLE 1. The DR program determines the rate of grid electricity and the SMP of June is taken as the rate of PV energy. Each EV connected to a nanogrid has a battery of 15 kWh capacity corresponding to 80% state-of-charge (SOC). EV charging is initiated with time-varying charging probability presented in Fig.3. The charging rate for each EV is 3kW and the EV charging efficiency is 90%. The initial SOC of EVs is set to 20% indicating deplete state of the battery. Since it takes 5 hrs to fully charge the battery, the EV charging taken in this work is considered slow charging. To fully charge EVs, the $d_{max}$ is set to 12hrs just like the $d_{max}$ for other flexible loads. For HVAC operation, the indoor temperature model and $CO_2$ density model in [62] are utilized. The target temperature is set to 23 degrees for the room where a resident is temporally located, and is set to 25 degrees for other rooms. Variation of outdoor temperature considered in simulations is shown in Fig.6(a). As shown in Fig.6(a), outdoor temperature is mostly above 23 degrees, requiring intensive power consumption of HVAC. The outdoor $CO_2$ density is set to 550 ppm and the target $CO_2$ density in rooms is set to 500 ppm.

A PV system representing rooftop PV panels and associated electronics of 3 houses is operated for a cluster. The PV system generating PV power following uniform distribution of the peak power over the range of 3~16kW is named "RPV #1" and the PV system producing PV power following uniform distribution of the peak power over the range of 2~11kW is named "RPV #2." The RPV #1 can produce more PV power on average and thus corresponding cluster can become a seller more often. This distinction of PV power production capacity might represent PV systems in remote communities with different specifications. Acting as a seller or a buyer for current time interval is determined by (2a) or (2b). Figure 6(b) shows temporal variation of PV power produced by RPV #1 and RPV #2. The black solid line and red solid line indicate the average variation of PV power production of RPV #1 and RPV #2, respectively. Considering typical variation of power consumption of a cluster consisting of electric loads in TABLE 1 and EVs, $PW^{max}$ is empirically set to 9kW for a cluster. To evaluate electricity cost, the DR program by the KEPCO and the SMP by the KPX [27] are used.

Daily PV power production curve of each cluster is obtained as follows. PV power production curves representing PV systems with different production capacities are presented in Fig.6(b). It is noted that PV power production curves shown in Fig.4(b) are obtained from a single PV system and PV power production curves in Fig.6(b) correspond to different PV systems with different production capacities. The PV power production curve shown as the lowest curve in Fig.6(b), which takes 2kW at 13hrs as peak production, is considered as the reference curve. The reference curve consists of values of PV power production at 10 minute time intervals. The reference curve takes value 0 before 6hrs and after 20hrs. Other PV power production curves taking different level of peak production are obtained by multiplication of power factor with the reference curve. For instance, the PV power production curve taking 6kW as peak production is obtained from the reference curve by multiplying the power factor 3. In each simulation involving PV power production, a PV power production curve is assigned to a cluster by assigning random peak power [66]. In simulations, peak production of clusters 1, 2, 3, 4, 5, 6 is 2.23kW, 5.87kW, 8.1kW, 10.34kW, 13.97kW, 16.0kW, respectively, when RPV #1 PV systems are used, and 1.72kW, 3.84kW, 5.47kW, 7.49kW, 9.01kW, 10.24kW, respectively, when RPV #2 PV systems are used.

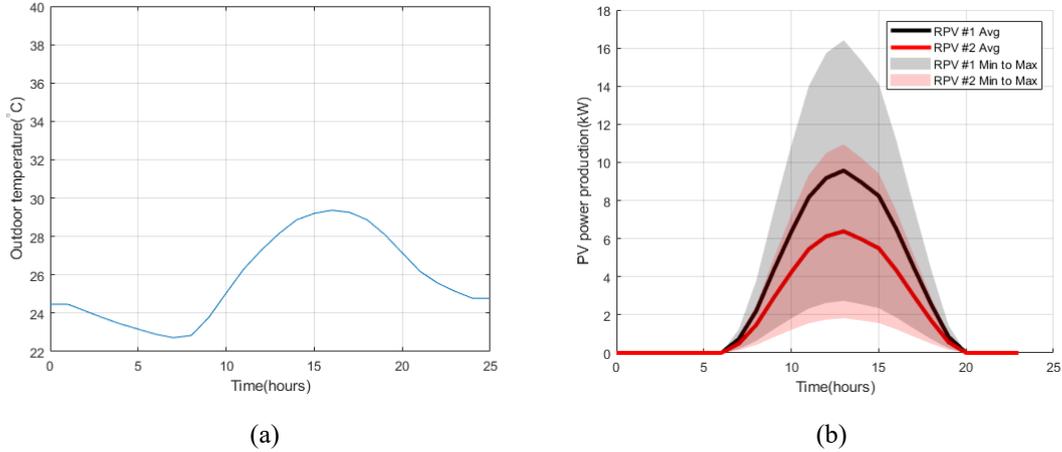

(a)                (b)

Fig.6. Variations of outdoor temperature and PV power production by RPV #1 and RPV #2 PV systems: (a) outdoor temperature; (b) PV power production. "RPV #1 Min to Max" for given time interval shows range of peak PV power production 3~16kW at 13hrs and "RPV #2 Min to Max" for given time interval shows range of peak PV power production 2~11kW. "RPV #1 avg" is the time-varying average of PV power production of RPV #1 and "RPV #2 avg" is the time-varying average of PV power production of RPV #2. It is noted that the color of the overlapped part in "RPV #1 Min to Max" and "RPV #2 Min to Max" becomes dark pink.

TABLE 2. Training parameters of GRU network

| Network training parameters | |
| --- | --- |
| Epoch | 1000 |
| Batch size | 200 |
| Learning rate | 0.005 |
| Gradient moving (average) | 0.9 |
| Dropout rate | 0.2 |
| Gradient threshold | 1 |

The GRU network is used to predict the future states of load demand and PV power production of each cluster. Training parameters of the GRU network shown in TABLE 2 are adopted from [67], [68], where similar architecture of the GRU network is studied. The total number of training epochs and batch size are set to 1000 and 200, respectively. The ADAM optimization algorithm is used for the GRU network with learning rate 0.005, gradient moving average 0.9, dropout rate 0.2, and gradient threshold 1 to update weighting coefficients such as $r_2$, $z_2$, $h_2$ in Fig.5(c) during training.

3 different scenarios are considered in simulations. The "w/o P2P trading" scheme represents power management without P2P trading. The "w/ P2P trading" scheme refers to conventional P2P trading, complying with trading condition in (2a). The "w/ proposed P2P trading" scheme indicates proposed P2P trading, complying with trading condition in (2b). Simulations are performed with the architecture of clusters presented in Fig.5(a). In each time interval, the role of each cluster is determined by (2a), (2b).

B. *Comparative Performance of Power Management based on Proposed P2P Trading*

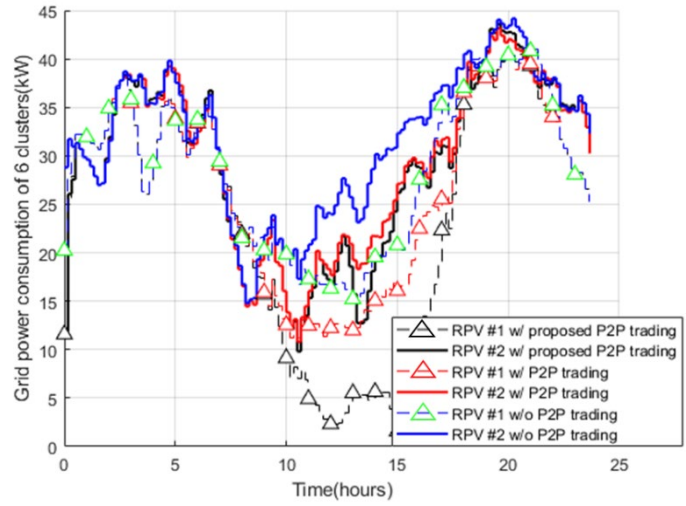

Fig.7. Variation of (total) grid power consumption according to 3 P2P trading schemes.

Variation of grid power consumption of 6 clusters according to power management based on 3 P2P trading schemes is presented in Fig.7. In general, RPV #1 PV system causes less grid power consumption than RPV #2 PV system, due to its larger PV power production on average. In case of "RPV #2 w/o P2P trading" scheme, the largest grid power consumption occurs due to the smaller PV power produced but unshared between clusters. With "RPV #1 w/o P2P trading" scheme, larger PV power production by RPV #1 PV system reduces grid power consumption to some extent. By the "RPV #1 w/ P2P trading" scheme, less grid power is consumed as compared to the "RPV #1 w/o P2P trading" scheme, because clusters trade excess PV power through P2P trading. Conventional P2P trading helps reduce the grid power consumption over the time zone 10-17hrs. In this time zone, power consumption of HVAC in the cluster, particularly the air-conditioners, is high due to outdoor temperature above 25degrees and thus takes the largest portion of grid power consumption. In this case, clusters might become buyers when satisfying the condition $PW_{load}(n) - PW_{PV}(n) > PW^{max}$. On the other hand, by the "RPV #1 w/

proposed P2P trading" scheme and "RPV #2 w/ proposed P2P trading" scheme, clusters that would be sellers in conventional P2P trading can be buyers if future states of load demand and PV power production require buying PV power. As a result, grid power consumption with proposed P2P trading is proactively reduced when the trend of grid power consumption rises. In this time zone, a large number of scheduled flexible loads are actually used with proposed P2P trading. When PV power production decreases beyond the time zone 10~17hrs, the power consumption of non-flexible loads such as HVAC is still held high due to high outdoor temperature. Therefore, grid power consumption is gradually increased regardless of power management schemes. The difference of grid power consumption between "RPV #2 w/ P2P trading" scheme and "RPV #2 w/ proposed P2P trading" scheme is less than the difference between "RPV #1 w/ P2P trading" scheme and "RPV #1 w/ proposed P2P trading" scheme, because the effect of P2P trading on grid power consumption is decreased due to reduction of PV power involved in P2P trading.

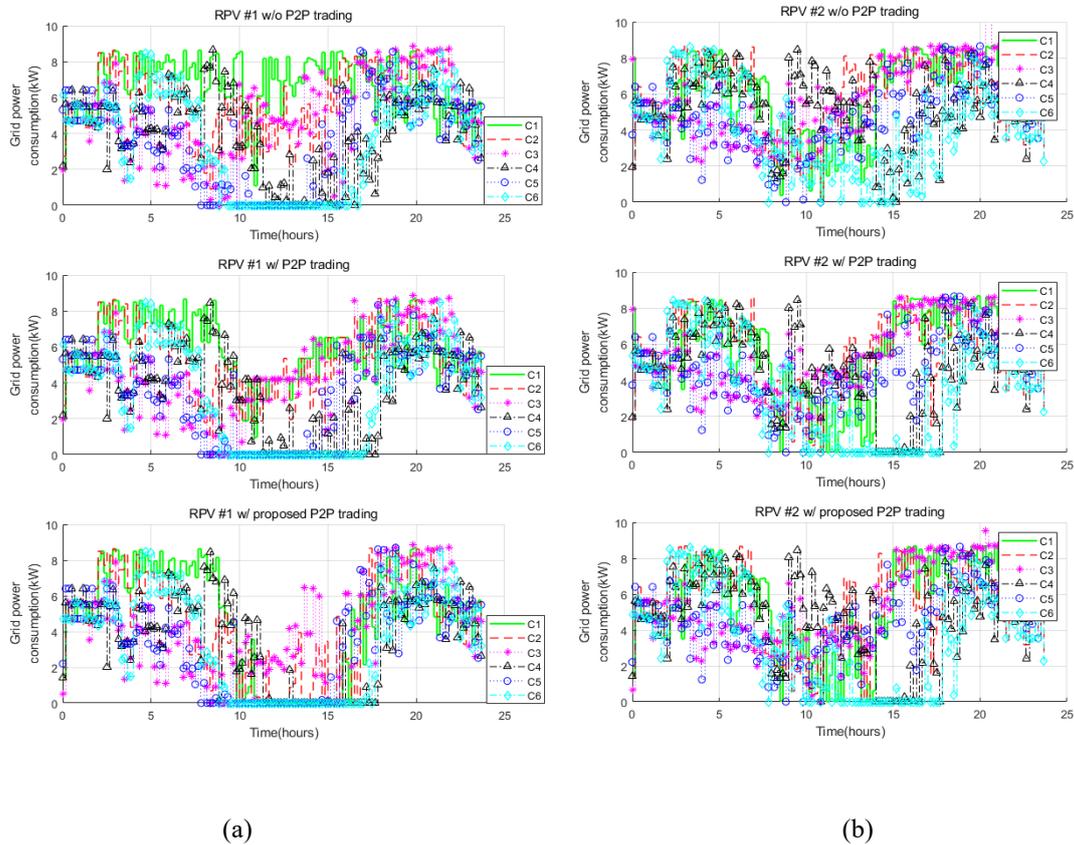

(a) (b)

Fig.8. Individual grid power consumption of 6 clusters with RPV #1 and RPV #2 PV systems: (a) with RPV #1 PV system; (b) with RPV #2 PV system.

Grid power consumption of each cluster is presented in Fig.8 to see the effect of P2P trading on the reduction of grid power consumption of each cluster. Figure 8(a) and Fig.8(b) are individual grid power consumption of 6 clusters with RPV #1 and RPV #2 PV systems, respectively, with and without P2P trading. Clusters are labeled according to the capacity of PV power production. Cluster 6 has the largest capacity of PV power production and cluster 5 has the second largest PV power production and therefore cluster 1 has the smallest capacity of PV power production. As a result, it is seen in Fig.8(a-b) that the cluster 6(C6) producing the largest amount of PV power consumes the smallest grid power on average over a day and the cluster 1 producing the smallest amount of PV power consumes the largest grid power on average over a day. In general, grid power consumption of each cluster is smaller with RPV #1 PV system, particularly over 10-15hrs when PV power production is relatively high, than with RPV #1 PV system.

In the case of the "RPV #1 w/o P2P trading" scheme, cluster 1 exhibits high grid dependency because it cannot buy PV power from other clusters even when self-supply PV power is insufficient for its own loads. On the other hand, cluster 6 is seen to have low grid dependency due to its large PV production capacity. With the "RPV #2 w/o P2P trading" scheme, grid dependency of clusters seems to be increased particularly over 10-15 hrs, as compared to the "RPV #1 w/o P2P trading" scheme because of their smaller capacity of PV power production. Grid dependency of clusters is seen to be decreased with P2P trading. The clusters 1,2 and 3 in the "RPV #1 w/ P2P trading" scheme indicate lower grid power consumption since the clusters trade PV power through P2P trading, unlike those with the "RPV #1 w/o P2P trading" scheme. Grid power consumption of clusters 1, 2, 3 with the "RPV #1 w/ proposed P2P trading" scheme seems to be decreased further by P2P trading based on the prediction of load demand and PV power production. Due to P2P trading by cooperative game strategy, the cluster 1 requires the most PV power through P2P trading and buys more PV power from other clusters than the clusters 2, 3. For the clusters 4, 5, 6, the difference of total grid power consumption between the "RPV #1 w/ proposed P2P trading" scheme and the "RPV #1 w/ P2P trading" scheme is small, because the amount of their PV power is sufficient for their load demand.

With RPV #2 PV systems, all clusters but cluster 6 exhibit high grid dependency due to lower capacities of PV systems. Due to smaller PV production capacity of RPV #2 PV system, on average, grid power consumption of 6 clusters increases in general. The "RPV #2 w/ P2P trading" scheme and the "RPV #2 w/ proposed P2P trading" scheme have small difference in grid power consumption of 6 clusters because of lower PV power traded by P2P trading.

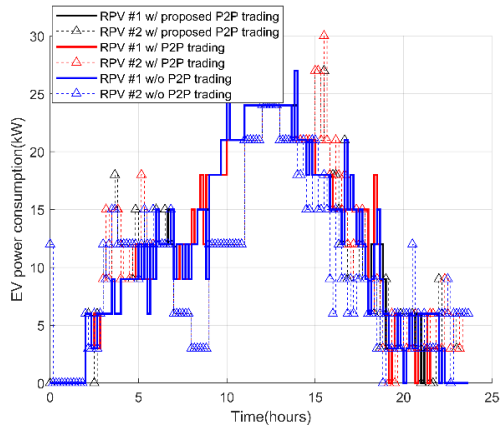

(a)

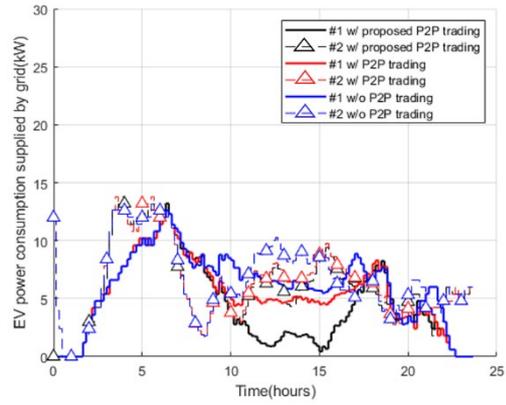

(b)

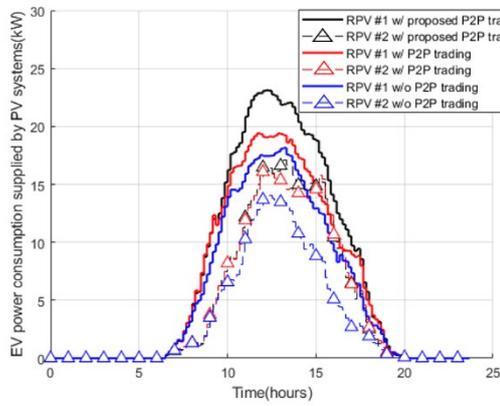

(c)

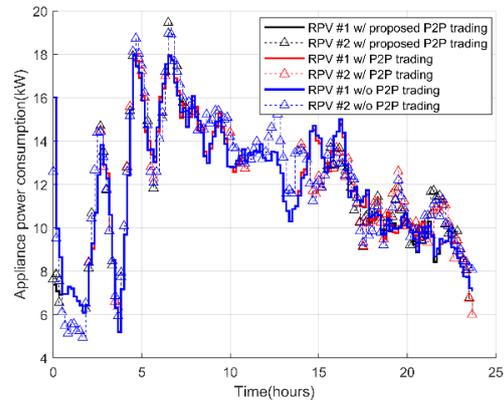

(d)

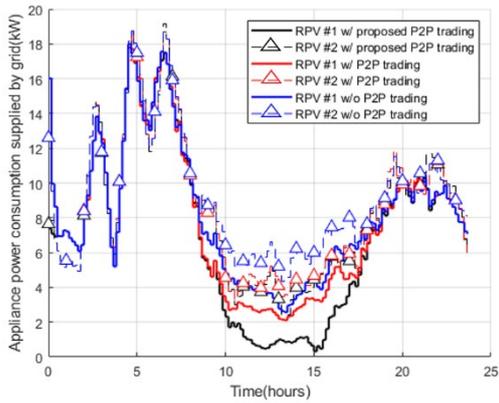

(e)

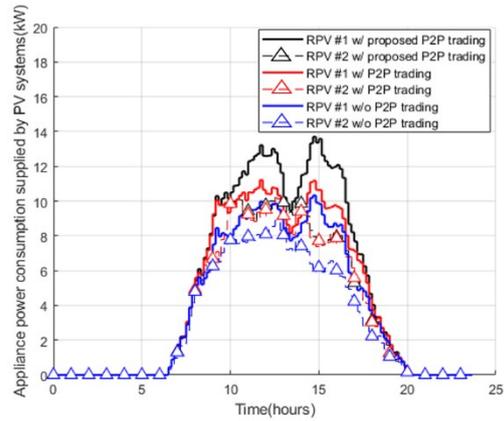

(f)

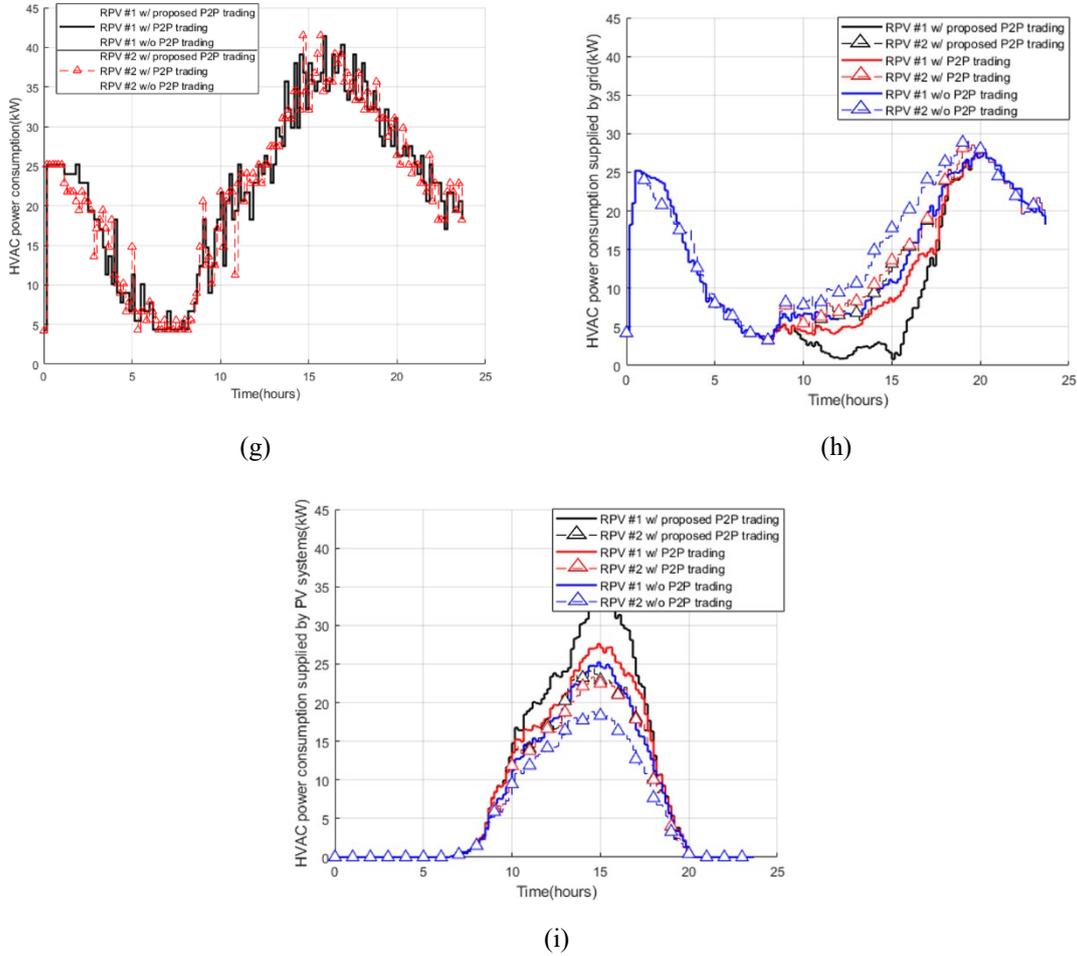

Fig.9. Power consumption of flexible loads and non-flexible loads in 6 clusters: (a) EV power consumption with 3 schemes; (b) EV power consumption supplied by grid with 3 schemes; (c) EV power consumption supplied by PV systems with 3 schemes; (d) power consumption of electric loads, excluding EVs and HVAC, with 3 schemes; (e) power consumption of electric loads, excluding EVs and HVAC, supplied by grid with 3 schemes; (f) power consumption of electric loads, excluding EVs and HVAC, supplied by PV systems with 3 schemes; (g) power consumption of HVAC with 3 schemes; (h) power consumption of HVAC supplied by grid with 3 schemes; (i) power consumption of HVAC supplied by PV systems with 3 schemes;

Figure 9 presents the temporal power consumption of 3 schemes with all loads in 6 clusters. The starting time of charging each EV follows the charging probability illustrated in Fig.3 and is common for 3 schemes. As seen in Fig.9(a), most EVs initiate charging from around 9hrs. All EVs are fully charged before 24hrs due to $d_{max}$=12hrs constraint. Two or more EVs are charged simultaneously around 12hrs, because of the relatively long charging time and relatively large PV power available. A major portion of EV power consumption comes from the grid before 7hrs and from PV systems during 7~17hrs, as shown in Fig.9(b-c). The Sum of power consumption supplied by grid and power consumption supplied by PV systems at each time interval is equal to EV power consumption. Large portion of power consumption of loads other than EVs and HVAC is supplied by grid before around 7hrs and supplied by PV systems during 7~17hrs, as shown in Fig.9(e-f). Power consumption of HVAC is supplied by grid before 7hrs and after 17hrs and supplied by PV systems during 7~17hrs. While power consumption of HVAC is high during 13~17hrs, power consumption of loads other than EVs and HVAC decreases due to $PW^{max}$ constraint. Since operation of non-flexible HVAC is determined by service plan depending on outdoor temperature, variation of power consumption of HVAC follows the trend of outdoor temperature variation, as shown in Fig.9(g).

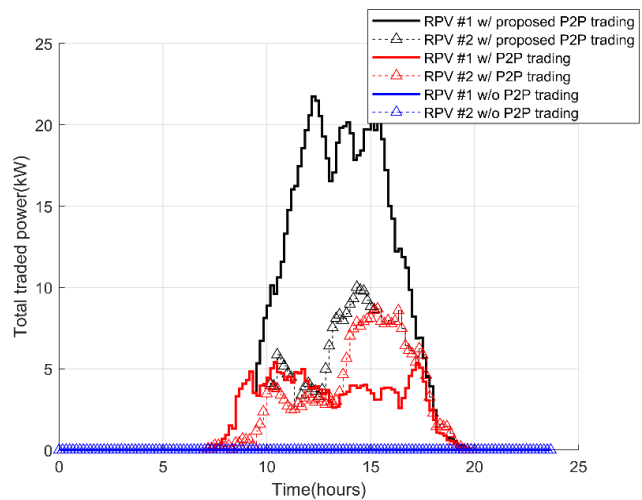

Fig.10. Total traded power with 3 schemes.

Figure 10 presents the total PV power transacted by P2P trading. The total traded power in the "RPV #1 w/ proposed P2P trading" scheme is larger than other cases because forecasted load demand is large and average PV production is higher, particularly over the time zone 10-15hrs. The total traded PV power in the "RPV #1 w/ P2P trading" scheme is lower because this scheme only considers current states of load demand and PV power production. The total traded power in the "RPV #2 w/ proposed P2P trading" scheme is still higher than the total traded power in the "RPV #1 w/ P2P trading" scheme in spite of smaller capacity of RPV #2 PV system. This means that clusters relying on proposed P2P trading method tends to consume PV power more than grid power. It is observed that in Fig.10 the "RPV #1 w/ proposed P2P trading" scheme utilizes more PV power around 12hrs when PV production reaches maximum in order to lower grid power dependency while the "RPV #2 w/ proposed P2P trading" scheme utilizes less PV power around 12hrs due to limited capacity of PV power production.

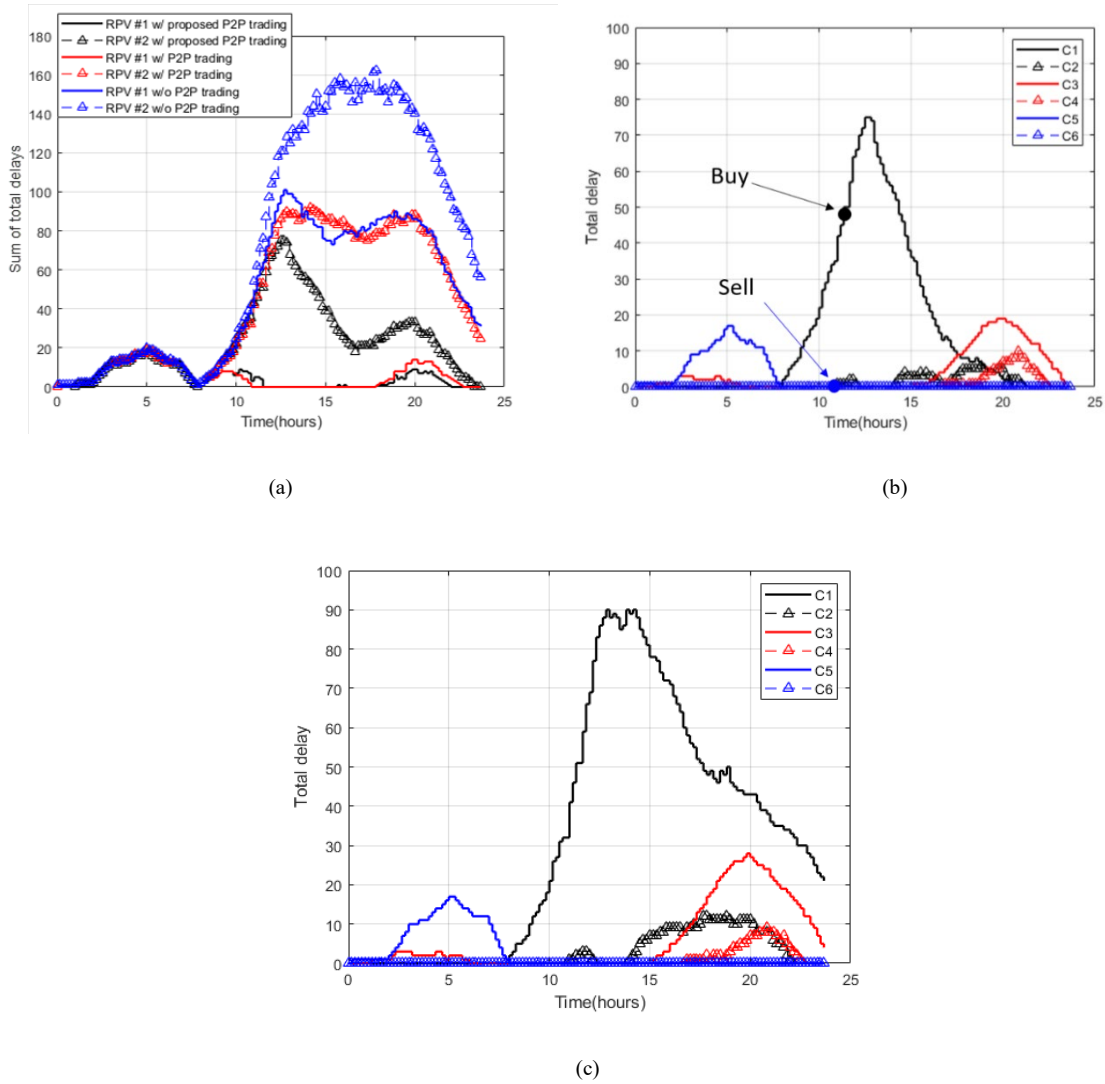

Fig.11. Sum of total delays of 6 clusters and total delay of each cluster: (a) total delay of flexible loads of 6 clusters in 3 schemes with RPV #1 and RPV #2 PV systems; (b) total delay of each cluster in "RPV #2 w/ proposed P2P trading" scheme; (c) total delay of each cluster in "RPV #2 w/ P2P trading" scheme. Unit of delay is 10 minutes.

Figure 11(a) shows the variation of the sum of total delays of 6 clusters according to 3 schemes. The total delay of flexible loads in a cluster is given by (5a). The "RPV #1 w/ proposed P2P trading" scheme exhibits the lowest sum of total delays thanks to plenty of PV power available for power management of clusters. Similarly, "RPV #1 w/ P2P trading" scheme shows a low sum of total delays because the sum of PV power production of 6 clusters is enough for maintaining a low sum of total delays. The Sum of total delays of "RPV #1 w/o P2P trading" scheme is close to the sum of total delays of "RPV #2 w/ P2P trading" scheme. Sum of total delays with the "RPV #2 w/o P2P trading" scheme is the largest at around 17hrs due to large power consumption of HVAC as seen in Fig.9(g). More PV power available for "RPV #1 w/o P2P trading" scheme at around 17hrs causes reduced sum of total delays of 6 clusters. A comparison of the "RPV #2 w/ proposed P2P trading" scheme with the "RPV #2 w/ P2P trading" scheme shows that the proposed P2P trading can reduce the sum of total delays of 6 clusters as a whole. The individual total delay of each cluster is presented in Fig.11(b-c). In Fig.11(b), exemplar time intervals of "buy" and "sell" PV power are presented. When the total delay of cluster 1 is large, the cluster buys the PV power to reduce the total delay, whereas, when the total delay of cluster 6 is zero, the cluster sells PV power. A similar variation of individual total delay is observed in Fig.11(c) with a different power management scheme.

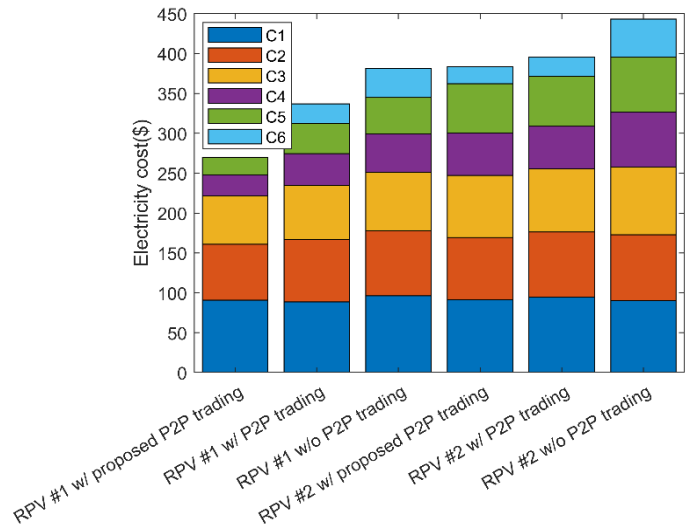

Fig.12. Comparison of electricity costs according to 3 schemes. Individual electricity costs of 6 clusters are marked by 6 different colors.

Individual electricity costs of 6 clusters are presented in Fig.12. The effect of the proposed P2P trading on the electricity cost paid by each cluster is visualized with different colors. The electricity cost of each cluster is calculated as the sum of the daily grid cost complying with the DR program and accumulated SMP in P2P trading over a day. The lowest electricity cost is charged to the "RPV #1 w/ proposed P2P trading" scheme. Particularly to cluster 6, electricity cost is not charged, due to income from selling large PV power through P2P trading can pay the grid electricity cost. A little electricity cost is imposed to the clusters 4, 5 because they sell PV power to other clusters in P2P trading. On the other hand, clusters 1, 2, 3 that often become buyers of PV power are associated with relatively large electricity cost. Comparing the "RPV #1 w/ proposed P2P trading" scheme with the "RPV #1 w/ P2P trading" scheme, the "RPV #1 w/ proposed P2P trading" scheme causes lower electricity cost thanks to more active P2P trading. The "RPV #1 w/o P2P trading" scheme represents the highest electricity cost among 3 schemes using RPV #1 PV system. The electricity cost of the "RPV #2 w/ proposed P2P trading" scheme is almost the same as that of the "RPV #2 w/ P2P trading" scheme because of smaller PV power traded in P2P trading. The "RPV #2 w/o P2P trading" scheme leads to the highest electricity cost due to smaller self-supply

PV power and unavailability of traded PV power. In the case of the electricity cost of "RPV #2 w/ proposed P2P trading" scheme, the electricity cost of cluster 4 is higher than that of cluster 5. This is due to the non-deterministic use of flexible loads with time-based rates of electricity.

TABLE 3. Comparison of electricity cost according to power management schemes.

| Scheme | Cluster index | Electricity cost ($) with RPV #1 | Electricity cost ($) with RPV #2 | Average electricity cost per cluster | Total electricity cost of 6 clusters |
|---|---|---|---|---|---|
| with proposed P2P trading | 1 | $90.59 | $91.20 | $45.02 (RPV #1)/ $63.81 (RPV #2) | $270.12 (RPV #1)/ $383.50 (RPV #2) |
| | 2 | $70.34 | $78.22 | | |
| | 3 | $60.68 | $77.83 | | |
| | 4 | $26.38 | $53.14 | | |
| | 5 | $21.82 | $61.72 | | |
| | 6 | $0.29 | $21.37 | | |
| with (conventional) P2P trading | 1 | $88.70 | $94.36 | $56.14 (RPV #1)/ $65.90 (RPV #2) | $336.88 (RPV #1)/ $395.42 (RPV #2) |
| | 2 | $77.99 | $81.86 | | |
| | 3 | $67.78 | $79.23 | | |
| | 4 | $39.94 | $53.22 | | |
| | 5 | $38.07 | $62.82 | | |
| | 6 | $24.38 | $23.90 | | |
| without P2P trading | 1 | $96.18 | $89.84 | $63.54 (RPV #1)/ $73.85 (RPV #2) | $381.50 (RPV #1)/ $443.14 (RPV #2) |
| | 2 | $81.52 | $83.07 | | |
| | 3 | $73.08 | $84.53 | | |
| | 4 | $48.32 | $69.20 | | |
| | 5 | $45.83 | $68.84 | | |
| | 6 | $36.29 | $47.63 | | |

In TABLE 3, a summary of the electricity cost incurred by different power management schemes is presented. This table is list version of Fig.12. The power management scheme with proposed P2P trading incurs the lowest electricity cost in general, followed by power management scheme with the conventional P2P trading and then power management scheme without P2P trading. By the proposed P2P trading method, total electricity cost of 6 clusters decreases by 29.2% and 13.5% with RPV #1 and RPV #2 PV systems, respectively as compared to those without P2P trading. It decreases by 19.8% and 3% in RPV #1 and RPV #2 PV systems, respectively as compared to the conventional P2P trading. For the cluster 6 with the largest PV power production capacity, the electricity cost with the proposed P2P trading decreases by 99.2% and 55.13% with RPV #1 and RPV #2 PV systems, respectively, as compared to those without P2P trading. It is notable in TABLE 3 that electricity cost($96.18) of cluster 1 with RPV #1 PV system is higher than the electricity cost($89.84) with RPV #2 PV system when P2P trading is not used for power management.

## V. CONCLUSION

In this paper, the power management of nanogrid cluster assisted by a novel peer-to-peer(P2P) electricity trading method is presented. The unbalance of power consumption among clusters is mitigated by P2P trading. For power management of each cluster consisting of 3 nanogrids, multi-objective optimization simultaneously minimizing total power consumption, grid dependency, and total delay incurred by scheduling is performed. Types of loads considered in this paper are flexible loads and non-flexible loads. The utility grid is taken as the primary source and the PV system is used as a secondary source. Self-supply PV power and traded PV power are used ahead of grid power. The temporal surplus of self-supply PV power of a cluster can be sold through P2P trading to other cluster(s) experiencing temporal shortage. The cluster in the temporal shortage of electric power buys the PV power and uses it to meet the load demand and reduce the total delay. In P2P trading, a cooperative game model is applied to maximize public welfare between buyers and sellers. To increase the efficiency of P2P trading measured by electricity cost, future trends of load demand and PV power production are taken into account by the proposed P2P trading. Unlike conventional P2P trading to resolve instantaneous unbalance between load demand and PV power production, the proposed P2P trading method attempts to do it by considering difference between trends of load demand and PV power production. For such purpose, the GRU network predicts the future load demand, representing power demand, and future PV power production, representing power supply. The effectiveness of the proposed P2P trading method for nanogrid clusters is verified by simulations, which demonstrate peak load reduction in peak hours, reduction of grid dependency, and total delay. By the proposed P2P trading, the total electricity cost of 6 clusters decreases by 29.2% and 13.5% with RPV #1 and RPV #2 PV systems, respectively, as compared to those without P2P trading. It decreases by 19.8% and 3% with RPV #1 and RPV #2 PV systems, respectively, as compared to the conventional P2P trading. For the cluster 6 with the largest PV power production capacity, the electricity cost with the proposed P2P trading decreases by 99.2% and 55.13% with RPV #1 and RPV #2 PV systems, respectively, as compared to the electricity cost with RPV #1 and RPV #2 PV systems without P2P trading. When the PV production capacity of clusters is large, i.e., RPV #1 PV system is used, significant reduction of grid power consumption of clusters is achieved by proposed P2P trading, as seen in Fig.7. The reduction of total delay of proposed P2P trading is more evident when the PV production capacity of clusters is relatively small, i.e., RPV #2 PV system is used, as witnessed in Fig.11(a).

## ACKNOWLEDGMENT

The authors gratefully acknowledge the support from the Energy Cloud R&D Program through the National Research Foundation of Korea (NRF) funded by the Ministry of Science, ICT (2019M3F2A1073314). Sangkeum Lee and Hojun Jin are co-first authors.